\documentclass{article}
\usepackage{PRIMEarxiv}
\usepackage[utf8]{inputenc} 
\usepackage[T1]{fontenc}    
\usepackage{hyperref}       
\usepackage{url}            
\usepackage{booktabs}       
\usepackage{amsfonts}       
\usepackage{nicefrac}       
\usepackage{microtype}      
\usepackage{lipsum}
\usepackage{fancyhdr}       
\usepackage{graphicx}       
\graphicspath{{media/}}     
\usepackage{multirow}
\usepackage{array}
\usepackage{amsmath}
\usepackage{xcolor}
\title{Real-time Sub-milliwatt Epilepsy Detection Implemented on a Spiking Neural Network Edge Inference Processor
\thanks{\textit{    
    Li and Zhao contributed equally to the work as first author.\\
    Corresponding author:
    wangdong@hainanu.edu.cn (D. Wang)
    \indent \indent yannan.xing@synsense.ai (Y. Xing)\\
    }: 
Published in Computers in Biology and Medicine, Volume 183, December 2024, 109225
DOI:https://doi.org/10.1016/j.compbiomed.2024.109225} 
}

\author{
  RuiXin Li, Guoxu Zhao, Dong Wang \\
  State Key Laboratory of Digital Medical Engineering\\
  Key Laboratory of Biomedical Engineering of Hainan Province\\
  School of Biomedical Engineering
  Hainan University\\
  Sanya\\
  China \\
  \texttt{wangdong@hainanu.edu.cn} \\
   \And
  Yannan Xing, Yuya Ling, Ning Qiao \\
  SynSense Co.Ltd. \\
  Chengdu \\
  China\\
  \texttt{yannan.xing@synsense.ai} \\
  \And
  Dylan Richard Muir, Karla Burelo, Mina Khoei \\
  SynSense AG \\
  Synsense, Thurgauerstrasse 60, Zürich \\
  Switzerland\\
}

\begin{document}
\maketitle

\begin{abstract}
Analyzing electroencephalogram (EEG) signals to detect the epileptic seizure status of a subject presents a challenge to existing technologies aimed at providing timely and efficient diagnosis. In this study, we aimed to detect interictal and ictal periods of epileptic seizures using a spiking neural network (SNN). Our proposed approach provides an online and real-time preliminary diagnosis of epileptic seizures and helps to detect possible pathological conditions.

To validate our approach, we conducted experiments using multiple datasets. We utilized a trained SNN to identify the presence of epileptic seizures and compared our results with those of related studies. The SNN model was deployed on Xylo, a digital SNN neuromorphic processor designed to process temporal signals. Xylo efficiently simulates spiking leaky integrate-and-fire neurons with exponential input synapses. 
Xylo has much lower energy requirments than traditional approaches to signal processing, making it an ideal platform for developing low-power seizure detection systems.
    
Our proposed method has a high test accuracy of 93.3\% and 92.9\% when classifying ictal and interictal periods. At the same time, the application has an average power consumption of 87.4 µW (IO power) + 287.9 µW (compute power) when deployed to Xylo. Our method demonstrates excellent low-latency performance when tested on multiple datasets. Our work provides a new solution for seizure detection, and it is expected to be widely used in portable and wearable devices in the future.
\end{abstract}


\section{Introduction}
\label{sec:introduction}
The measurement of electroencephalogram (EEG) is a safe and non-invasive method for recording brain signals by attaching electrodes on the scalp \cite{1.1.1}, where the continuously recorded data can reflect the electrical activity of neurons and the electrical potential across the entire brain surface. As the characteristic patterns of EEG signals are distinguishable at different neural states, brain activities can be analyzed and studied using EEG \cite{1.1.2}.
        
        Epilepsy is a common neurological disorder characterized by recurrent episodes of abnormal brain discharges, behavioral seizures, or abnormalities in sensation, emotion or consciousness \cite{1.1.3}.
        Currently, there are more than 50 million epileptic patients in the worldwide. Thus, the diagnosis and treatment of epilepsy are of great social and economic value. By detecting the EEG signals, the different degrees of abnormal discharge in epileptic patients can be diagnosed and monitored. This can help to determine the type and location of epilepsy and to choose appropriate treatment options during epilepsy diagnosis, and help to monitor treatment effect, adjust treatment plan timely and avoid unnecessary drug side effects during epilepsy treatment. In recent years, with the development of computer and artificial intelligence (AI) technologies, EEG detection has been widely used in epilepsy researches. By analyzing and mining a large amount of EEG data, it becomes realizable to explore the pathophysiological characteristics of epilepsy and develop more accurate methods for the diagnosis and treatment of epilepsy \cite{1.1.4,1.1.5}.
    
        Real-time EEG monitoring plays a crucial role in medical diagnosis by providing essential information about the health status of human bodies \cite{1.3.1,1.3.2}.
        This can aid doctors in timely disease detection and is vital for formulating and adjusting treatment plans.
        Moreover, real-time monitoring of biological signals is essential in scientific research fields, such as neuroscience.
        Researchers can investigate the relationship between specific behaviors or cognitive tasks and brain activity by using EEG signals.
        Given the time-consuming and labor-intensive nature of manual detection, an artificial intelligence model is urgently required to perform real-time monitoring while accurately identifying abnormal EEG signals. 
        
       Artificial neural networks (ANNs) are now widely used for solving problems in signal analysis, and have been widely employed in numerous fields, such as speech processing, computer vision and natural language processing. \cite{1.2.2}.
        However, there are significant differences between traditional ANNs and real biological neural networks.
        Traditional ANNs have inputs and outputs that are both real numbers, while the information transmission in the human brain is in the form of discrete action potentials or spikes.
        Spiking neural network (SNN) is the third generation of neural networks, and models the behavior of biological neurons, where information is transmitted in the form of discrete electrical impulses, known as spikes.
        SNNs have shown promise in accurately identifying patterns in time-series data, such as EEG signals, due to their ability to efficiently handle temporal information.
        Our research focuses on real-time detection of EEG signals using low-power neuromorphic chips which deploy spiking neural networks for inference.
        These chips can quickly and efficiently detect epileptic signals, and trigger an appropriate response.
        
        Our study utilized the CHB-MIT dataset from Boston Children's Hospital \cite{chb-mit} and the Siena Scalp EEG dataset from the University of Siena \cite{detti2020siena}.
        The Wavesense neural network model \cite{5.2.1wavesense} was combined with time-series analysis and Delta-Sigma coding techniques for detecting and identifying epilepsy signals based on spikes characteristics. The feasibility and accuracy of this method were further verified by experiments on Xylo, a low-power neuromorphic processor for signal processing inference. Our experimental results demonstrate an ultra-low power consumption and high accuracy when deployed to the Xylo processor, suggesting that the great promise of this SNN-based epilepsy identification method for serving as a new epilepsy diagnostic tool in the future. Our work can enhance the efficiency of epilepsy diagnosis, reduce the power consumption during signal acquisition, and provide guidance for the practical application and dissemination of spiking neural networks.

\section{Related Work}
\label{sec:related_work}
Historically EEG signal identification has been performed manually, which is a time-consuming and subjective process, and susceptible to human error.
    Although linear signal processing techniques such as time-domain, frequency-domain, and time-frequency analysis have become popular ~\cite{2.1.1}, they fail to accurately capture the nonlinear characteristics of complex electrical activity in the brain.
    However, with the development of nonlinear EEG data classification methods, artificial neural networks have become an essential tool for nonlinear analysis and identification of EEG signals.
    Various neural network models have been explored, including spatiotemporal convolutional neural networks \cite{scnn}, fuzzy function-based classifiers \cite{ff}, and long short-term memory recurrent neural networks \cite{2018long}.
    Traditional ANNs are highly dependent on feature extraction, and the quality of feature extraction can greatly affect their classification accuracy. Therefore, past researchers have spent a lot of time and effort developing suitable feature extraction techniques.
    In contrast,  SNN models show potential for processing complex spatiotemporal patterns without the need for manual feature extraction. 
      
     As early as 2007, Ghosh Dastidar developed an efficient SNN model for classification and epilepsy detection, which uses RProp as the training algorithm and achieved a classification accuracy of 92.5$\%$ \cite{2.2007improved}.
     In 2009, Adeli developed a new multi-spike neural network (MuSpiNN) model and a new supervised learning algorithm called Multi-SpikeProp, which is used to train MuSpiNN, which can achieve complex EEG classification problem achieves a classification accuracy of 90.7$\%$ to 94.8$\%$ \cite{2.2009new}.
     The creation of the NeuCube framework in 2014 by Kasabov et al, based on SNN, marked the world's first development environment for building brain-inspired artificial intelligence. Studies have shown that the NeuCube model can improve the accuracy of brain spatiotemporal data classification compared to standard machine learning techniques \cite{2.2014neucube}. In 2022, Karla et al. detected interictal high-frequency oscillations (HFO) in scalp EEG using a spiking neural network. The presence of HFOs is associated with active epilepsy and has an accuracy rate of 80\%. However, it should be noted that HFOs are not a mandatory requirement for epileptic seizures, and the network was not implemented on a neuromorphic chip \cite{karla}.

    If SNN-based EEG signal analysis is to be applied clinically, the SNN network must be deployed to neuromorphic inference hardware.
    Currently, there are few studies on epilepsy detection based on spiking neural networks that can be implemented on neuromorphic chips.                                 

\section{Spiking Neural Network}
\label{sec:spiking_neural_network}
\begin{figure}[htbp]
    \begin{center}
    \includegraphics[width=4cm,scale=.75]{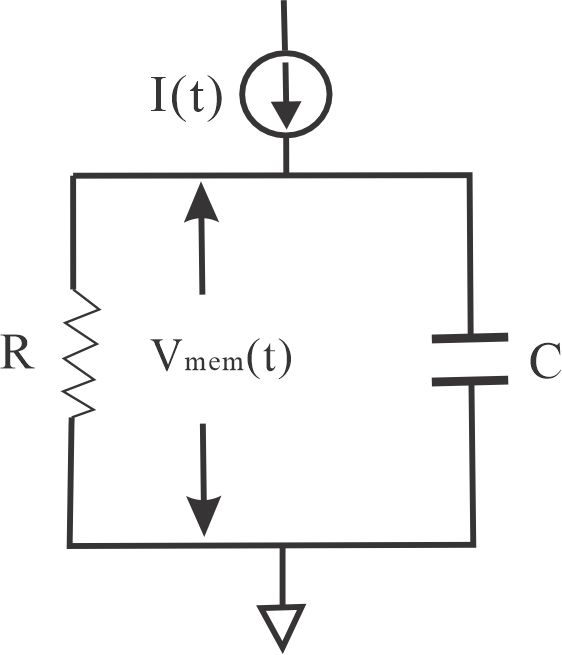}
    \end{center}
    \caption{Equivalent circuit of the LIF neuron model.}\label{curcuit1}
    \end{figure}

\begin{figure}[h]
    \begin{center}
    \centerline{\includegraphics[width=1\linewidth]{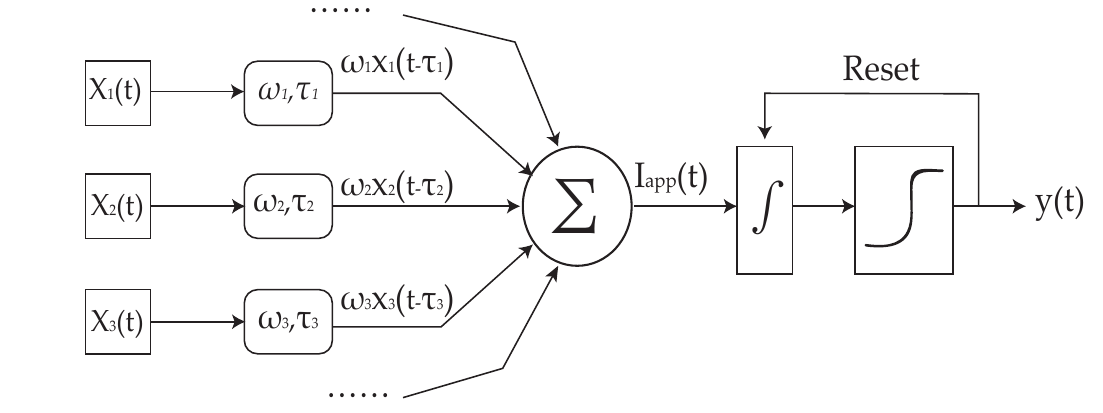}}
    \end{center}
    \caption{ Block diagram of a Leaky integration and Fire spiking neuron.}\label{fig-LIF}
    \end{figure}

Spiking Neurons are a mathematical construct inspired by the dynamics and behaviour of biological neurons.
Spiking Neurons receive inputs and communicate with other neurons through discrete binary events known as Spikes.
Neurons temporally integrate these signals in small dynamical systems representing synapses and neuron membranes, and when a membrane state passes a configurable threshold, emit a Spike to communicate with other connected neurons.
Similar to an ANN, layers of spiking neurons are connected to each other through linear weights, which effectively scale the strength of inputs received by synapses \cite{w1}.
Here we describe an LIF (Leaky integration and Fire) neuron~\cite{1.2.1LIF}, one of the simplest spiking neuron models. 
    The state of the neuron, known as the membrane potential $V_m(t)$, evolves over time depending on its previous state as well as its inputs \cite{1.2.5}.The equivalent circuit of a LIF neuron is shown in the Figure. \ref{curcuit1}.
    The dynamics of this state can be described as follows Equation \ref{e1}.
    \begin{equation}\label{e1}
        RC\frac{dV_{\text{mem}(t)}}{dt}=-V_{\text{mem}}(t)+I(t)R
    \end{equation}
     In the equation, $RC\frac{dV_{\text{mem}(t)}}{dt}$ represents the rate of voltage change across the capacitor, $-V_{\text{mem}}(t)$ represents the decay of the membrane potential due to the leakage resistance, and $I(t)R$ represents the voltage change caused by the input current. Overall, this equation reflects how the membrane potential increases when the neuron receives input current and decreases due to the leakage effect when there is no input current.The working mechanism of the LIF neuron model is: when the membrane potential $V_{\text{mem}}(t)$ exceeds a certain threshold, the neuron generates a "spike" (i.e., "fires"), and then the membrane potential is reset to a lower value, simulating the discharge process of a biological neuron.
      Figure \ref{fig-LIF} provides a detailed illustration of this process. 
      The input current $I_{app}(t)$ is the sum of weighted input signals filtered by a set of synapses, where each input $x(t)$ is weighted independently by $\omega$, and can be positive or negative, and subject to synaptic filtering with time constant $\tau$.
    The result is a spatially summed time series.
    The input current $I_{app} (t)$ travels to the neuron \textit{soma}, which acts as a low-pass filter and integrates input over time, updating the internal state variable $V_m$.
    The soma performs integration and applies a threshold to make a decision on whether to spike or not.
    After a spike is produced, the voltage $V_m$ is reset to a value $V_{reset}$.
    Finally, the resulting spike is transmitted to the other neurons in the network, this network consisting of many similar LIF neurons is called a SNN.
SNNs can adopt a wide range of network architectures, similar to standard ANNs.
Neurons in different layers are connected through synapses with multiple dynamically adjustable weights, which transmit signals from the input layer to the next layer.
Different topological structures have different characteristics.
Feedforward SNNs are composed of an input layer, one or more hidden layers, and an output layer, with dense connections passing between each layer in a strictly forward manner\cite{1.2.6feed}.
Recurrent SNNs are also common, raising additional complexities with regard to stability and learning algorithms\cite{1.2.7demin2018recurrent}.
Architectures with lateral inhibition (``Winner-Take-All'' networks) are used for SNNs\cite{1.2.8cyclic}, with reference to concepts from Neuroscience\cite{3.e}.
In this work we choose feedforward SNNs, due to their simplicity and ease of training.

\section{Xylo neuromorphic processor}
\subsection{Architecture and Functional capabilities}   
\begin{figure}[htbp]
\begin{center}
\includegraphics[width=8cm,scale=.75]{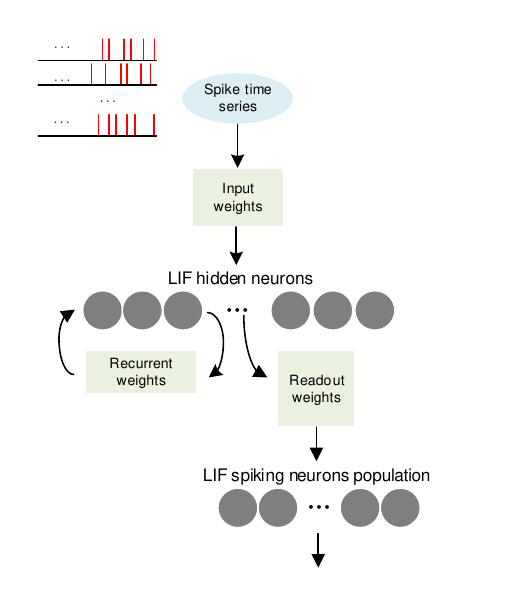}
\end{center}
\caption{Xylo comprises a hidden population of 1000 LIF neurons and a readout population of 8 LIF neurons. Dense input and output weights are available, along with sparse recurrent weights that can target up to 32 hidden neurons per neuron. The inputs, which consist of 16 channels, and the outputs, which consist of 8 channels, are transmitted through asynchronous firing events.}\label{fig-xylo}
\end{figure}

Xylo is an application-specific integrated circuit (ASIC) that utilizes an all-digital approach for simulating spiking leaky integrate-and-fire neurons with exponential input synapses.
It is designed to be energy efficient and highly configurable, with the ability to adjust synaptic and membrane time-constants, thresholds, and biases for individual neurons.
Xylo supports a wide range of network architectures, including recurrent networks, residual spiking networks, and other arbitrary configurations.
The overall logical architecture of Xylo is illustrated in Figure \ref{fig-xylo}.
It has up to 1000 digital LIF hidden neurons with independently configurable time constants and thresholds, and each neuron supports up to 31 spikes generated per time-step.
Xylo also has 8-bit input, recurrent, and readout weights with bit-shift decay on synaptic and membrane potentials.
It supports one output alias per hidden neuron, one input synapse, and one output spike per time-step in the readout layer.
Furthermore, the Xylo ASIC allows for various clock frequencies, and the network time step $dt$ can be chosen freely.
Overall, Xylo is a flexible and powerful architecture suitable for many applications 
 \cite{xylo}.

To minimize power consumption and improve performance and reliability, the Xylo chip incorporates a low-power digital circuit design. This chip features sparse recurrent weights, with a maximum of 32 targets per hidden neuron, reducing energy and computation by connecting only a few neurons. Compared to dense connections, sparse connections also decrease memory bandwidth requirements, computation time, and power consumption.To measure the aforementioned power consumption, the Xylo development kit seamlessly integrates an onboard current monitor, allowing users to sample real-time current data at a frequency of 1000Hz via an ADC (Analog-to-Digital Converter). The kit includes two distinct power tracks: "core" and "IO." The core power track covers the total power consumption of the Xylo frontend and processing core, including RAM read/write operations, logic circuit operations, event routing, and other essential processes. Conversely, the IO power track pertains to the chip interface power, primarily facilitating Serial Peripheral Interface (SPI) operations.

The Xylo chip employs an off-chip training and on-chip inference approach to learn data samples. Off-chip training refers to the phase where we use an external computing platform to train our neural network model. During this phase, the model learns features and weights through a large amount of labeled data, often involving the backpropagation algorithm to optimize network parameters. Once the model is trained and the parameters are set, we convert these parameters (weights, biases, thresholds, etc.) into a format suitable for the Xylo chip. This may include quantization and encoding of the parameters to fit the hardware architecture of Xylo. In the on-chip inference phase, the converted model parameters are loaded onto the Xylo chip. The Xylo chip utilizes its digital spiking neural network to perform real-time event-driven inference. Designed for energy efficiency and real-time processing, the Xylo chip is capable of executing complex signal processing tasks with very low power consumption.

\subsection{Mapping and quantization}
For deploying SNNs in Rockpool \cite{3.2rock}, the structure of an SNN is converted to a computational graph.
Nodes in the computational graph contain parameters of a neuron model, such as thresholds, biases, etc.; high-level structures such as dense weights; as well as representing the connections between individual layers in the network.
When a SNN is deployed to a Neuromorphic Chip, the topology of the neural network needs to be mapped to the actual physical layout of the chip.
This enables the implementation of the same computational process as the original SNN on the chip.
The mapping process converts the neuron graph to a form which matches the hardware architecture.
In order to build the hardware-equivalent configuration, neuron IDs, weights, inputs and outputs, and other required information are extracted from the graph.
In addition, when performing calculations on Xylo, it is necessary to quantise floating-point parameters in the trained SNN to low-precision integer values.
To convert weights and thresholds, one needs to find the absolute maximum of all input weights to a neuron, and calculate a scaling factor such that this value can be mapped to a range of ±128, and the threshold is scaled by the same scaling factor.
After scaling, weights and thresholds are rounded to the nearest integer, which is the process of quantization.

\section{Experiments}
\subsection{Dataset}
\subsubsection{Siena scalp EEG database}
The first dataset used in this work is the Scalp Electroencephalography Database of the Department of Neurology and Neurophysiology at the University of Siena, Italy \cite{detti2020siena}.
This database contains EEG records from 14 patients, including 8 males (aged 25--71) and 6 females (aged 20--58).
This experiment used Video-EEG monitoring under a sampling rate of 512~Hz, and the electrode arrangement followed the standard-1020 system.
Most records also include 1--2 electrocardiogram signals.
According to the standards of the International League Against Epilepsy, clinical doctors carefully checked the clinical and electrophysiological data of each patient and made a diagnosis and classification of epilepsy.
In these 14 patients' records, EEG signals were recorded before and after 1--10 seizures, and the recording time ranged from 145--1408 minutes.

\subsubsection{CHB-MIT scalp EEG database}
The second dataset we used was collected by Boston Children's Hospital and contains 23 cases involving 22 children with intractable epilepsy \cite{chb-mit, 2CHB-MIT}.
These patients recorded their epileptic seizures after discontinuing antiepileptic drugs and undergoing monitoring for several days to evaluate the potential for surgical intervention.
The subjects in this database include 5 males and 17 females, aged between 1.5 and 22 years old.
Cases chb21 and chb01 are data from the same female subject, with a gap of 1.5 years between them.

\subsection{Preprocessing}
\begin{figure*}[htbp]
\begin{center}
\includegraphics[width=\textwidth]{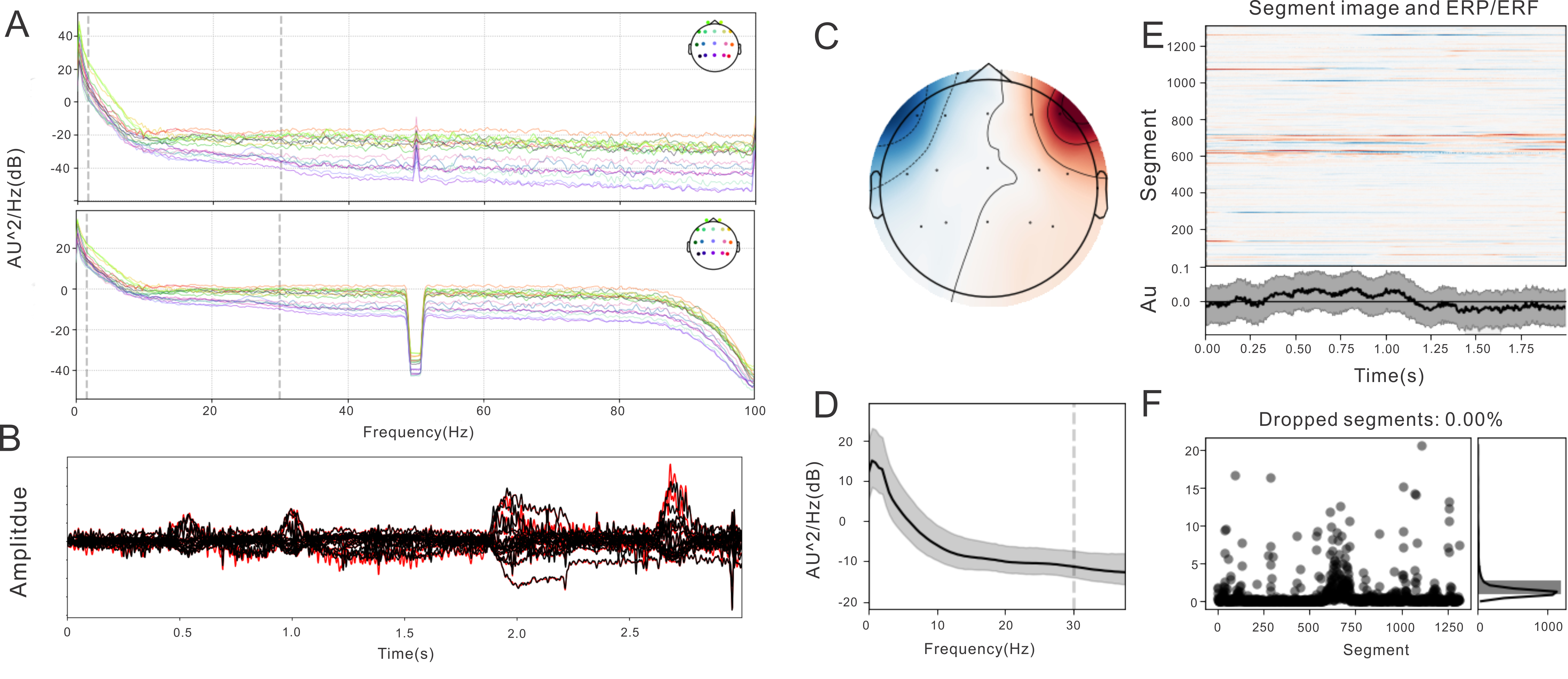}
\end{center}
\caption{\textbf{(A)}. Comparison of power spectrum before and after filtering and notching. \textbf{(B)}. EEG comparison after removing the selected independent components. Red is the signal before ICA, and black is the signal after ICA. This operation can check the influence of a specific component on the overall signal after removal. \textbf{(C)}. Spatial distribution of each component on the scalp surface, the brighter the area, the greater the contribution of the independent component to the area. \textbf{(D)}. The energy distribution of this independent component at different frequencies. \textbf{(E)}. The time series reconstructing the trails of the independent components and the time of each trail. \textbf{(F)}. The total variance of the waveform of each component over time under different numbers of independent components over time.}\label{perprocess}
\end{figure*}
We begin by describing the preprocessing of the Siena Scalp EEG dataset.
In this dataset, the arrangement of electrodes is placed according to the standard10-20 system.
However, the order of electrodes differs between subjects.
In addition, the researchers also captured ECG signals from some subjects that were unrelated to the experiment.
The common electrodes of all subjects in the standard 10-20 system are kept, and the order of all electrodes is uniformly sorted.
After these operations, the electrode positioning is performed on the original signal once, and the electrode reference point is reselected.
The data is re-referenced according to the average value of all signals using this function.
Subsequently, the influence of the acquisition environment, such as power frequency interference in the acquisition system, needs to be eliminated.
In addition, studies have shown that EEG signals only contain valid information at 1--80 Hz.
Due to the need for a large number of repetitive preprocessing operations on the data in the study, the EEG data of all patients were processed with 1--80Hz band-pass filtering, and then 50Hz notch processing to eliminate power frequency interference.
In order to verify the data processing effect of filtering and notching, We show the comparison effect of power spectral density maps before and after filtering Figure \ref{perprocess}A.

Independent components analysis (ICA) is widely used to remove ECG, eye movement, myoelectricity and head movement signals in EEG, and the effect is remarkable \cite{ica1}.
ICA is a method of linear transformation based on statistical principles, which separates data or signals into linear combinations of statistically independent non-Gaussian signal sources \cite{ica2, ica3}.
Since the collected signal is a mixed signal of spontaneous EEG signal and various noises, it meets the conditions of use of the ICA algorithm.
After observing the effect before and after ICA Figure \ref{perprocess}B, the appropriate independent components were selected to be removed.
\begin{figure*}[htbp]
\begin{center}
\includegraphics[width=17cm,scale=.75]{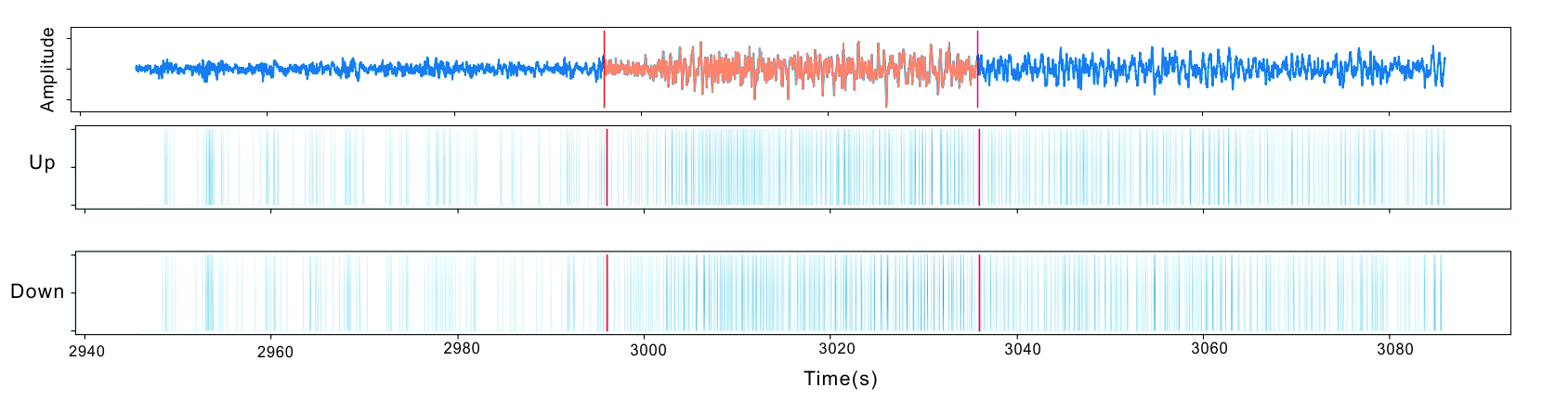}
\end{center}
\caption{Original signal and corresponding spikes time series, the data in the red line is the data of the epileptic seizure period.}\label{ds}
\end{figure*}
Figure ref{perprocess}C, D, E, F shows the independent component information of ICA. the four graphs show the spatial distribution of each component on the scalp surface (the brighter the area, the greater the contribution of the independent component to the area), the energy distribution of this independent component at different frequencies, the time series reconstructing the trails of the independent components and the time of each trail, and the total variance of the waveform of each component over time under different numbers of independent components over time \cite{ica4}.

In addition to the above operations, operations such as resampling and baseline correction are also involved, which are not meaningful to describe here. In the CHB-MIT dataset, our preprocessing pipeline closely resembles the preprocessing pipeline used by the Siena Scalp EEG Database. However, the electrodes used in this dataset vary across subjects and time periods, which results in inconsistent matrix dimensions when fed into the pulse neural network. therefore, we only applied the aforementioned preprocessing steps on the C3-P3 and C4-P4 electrode channels, which exhibited the most pronounced epilepsy symptoms.
The preprocessed signal is then converted to a spiking time series using a sigma-delta encoder\cite{ica5},  this encoder converts the original signal into spikes signals by partitioning the signal's amplitude range into multiple equidistant intervals. When the signal crosses one of these intervals, it generates either an up or down type spike signal based on the polarity of the signal's slope at that point.
The result of this encoding in shown in Figure \ref{ds}.
The highlighted portion indicates the epileptic phase.
Since the frequency and amplitude of the original signal increases during the patient's seizure, so does the number of spikes.
In this method of encoding, information from individual channels of EEG data will be transformed into two separate time series of spiking events. These two series correspond to the upward and downward trends in the EEG signal, respectively.

 \subsection{Network}
\begin{figure}[htbp]
    \begin{center}
    \includegraphics[width=8cm]{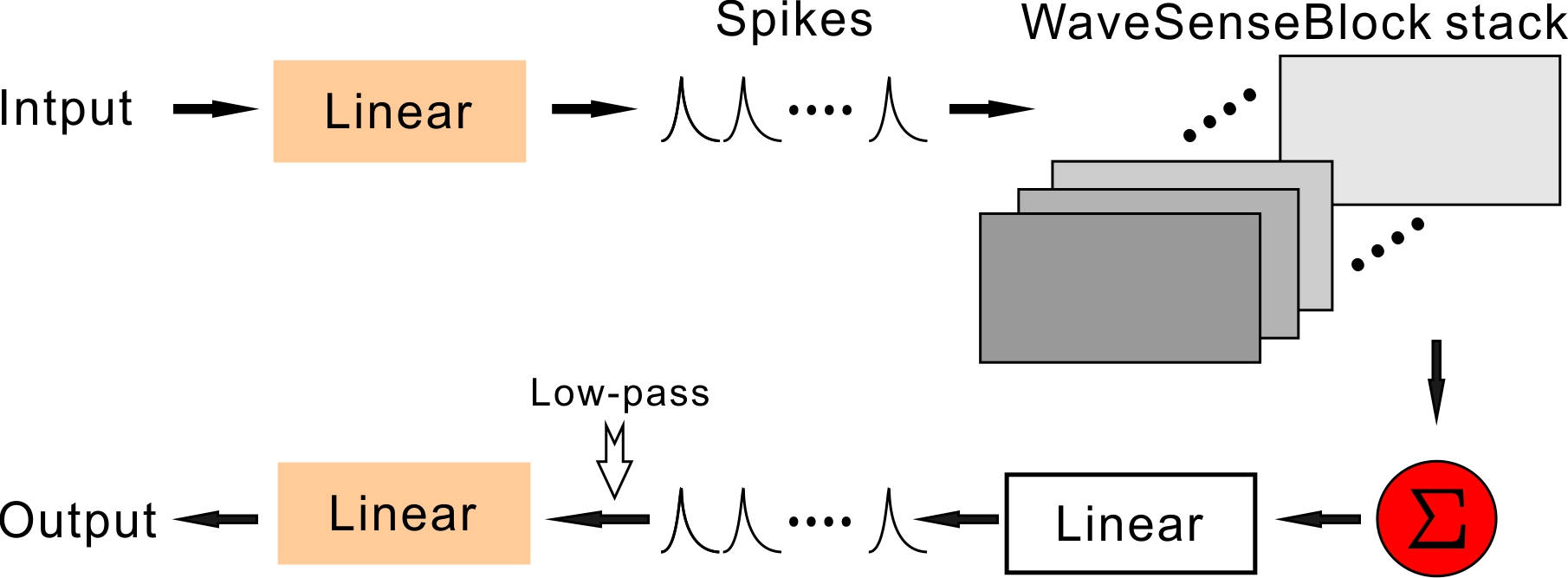}
    \end{center}
    \caption{Overview of the entire architecture.}\label{wave}
    \end{figure}

    During the training process, the employed model is WaveSense \cite{5.2.1wavesense}.
    This model takes spike time series data as input and is constructed using a network architecture based on spiking neural networks.
    It draws inspiration from the architecture of WaveNet \cite{5.2.2wavenet}.
    The model adopts a stacked network architecture, which stacks multiple blocks together to gradually extract higher-level features of the input signals.
    Each block consists of multiple computational modules, including dilated temporal convolutional layers, gated convolutional layers, and pooling layers.
    The dilated temporal convolutional layer uses multiple synapse projections with different time constants to achieve dilated temporal convolution.
    The gated convolutional layer uses gate mechanisms to regulate information flow, and the pooling layer is used to reduce the dimensionality of feature maps.
    These computational modules are used to extract the temporal features of the input signal and transform them into classification or regression results.
    The WaveSense model achieves audio and signal processing tasks by converting the input signal into a pulse sequence and processing it with a series of pulse neural layers.
    The model has been tested on multiple datasets and has achieved excellent performance.
    Advantages of the network include its strong generalization ability and robustness, suitability for various low-dimensional signal processing tasks, and ability to be implemented on neuromorphic hardware with low power consumption and high efficiency.
    The WaveSense model can be trained using the backpropagation algorithm and compared to other deep learning models.
    In this experiment, we set the number of neurons in the hidden layer of the readout to 16, the number of synapses in the dilated layer of the block to 2, and the membrane potential time constant of neurons to 0.002. The threshold for firing spikes of all neurons was set to 0.6.
    Figure. \ref{wave} shows the overview of the model architecture, including data processing, spike encoding, stacked network architecture, and output layer, to convert low-dimensional signals into category probability distributions.
    \subsection{Training and deploying}
    While experiencing reduced average latency, experiments demonstrate that optimal model prediction accuracy is achieved by segmenting the signal into 5-second samples. Consequently, the initial dataset undergoes partitioning into numerous 5-second trials, each accompanied by a corresponding label.
    These trials are then processed to create spike time series data, which is fed into the network.
    The processed samples are randomly split into training and testing datasets at a 4:1 ratio, trained for 150 training epochs, a learning rate of 0.0005, and an output dimensionality of 2.
    Back Propagation Through Time (BPTT) \cite{5.2.2BPTT} is used to train the SNN, and the peak current of the neurons on each segment is calculated by extracting the synaptic current of the output layer.
    The prediction is made by choosing the neuron with the highest peak current, and cross-entropy loss \cite{5.2.3cross} is calculated with the label to obtain $L_{CE}$. The optimization technique employed is Adam\cite{adam}.
    It is noteworthy that the model is intended to be used in streaming mode, which necessitates an appropriate loss function.
    The activity of LIF neurons may change substantially during learning, resulting in either a lack of spikes or excessive energy consumption in neuromorphic implementations \cite{5.2.4sorbaro2020optimizing}.
    To maintain sparse activity and limit the activity of these neurons, an activity regularizer term is included in the loss function.
    The final loss function is given by Equation \ref{loss}:
    \begin{equation}\label{loss}
        L=L_{CE}+ \sum\sum_{i} \left(\frac{N_i^t\Theta(N_i^t-l)}{T\cdot N_{neurons}}\right)^2
    \end{equation}
    The activation loss is determined by the network's population size $N_{neurons}$ and the total number of excess spikes produced in response to an input of duration $T$ time steps.
    The excess spikes are those that exceed a certain threshold $l$, and are summed over all neurons $N_i$ and time bins $t$. where $\Theta$ is the Heaviside step function.

\begin{figure}[htbp]
    \begin{center}
    \includegraphics[width=9cm,scale=.75]{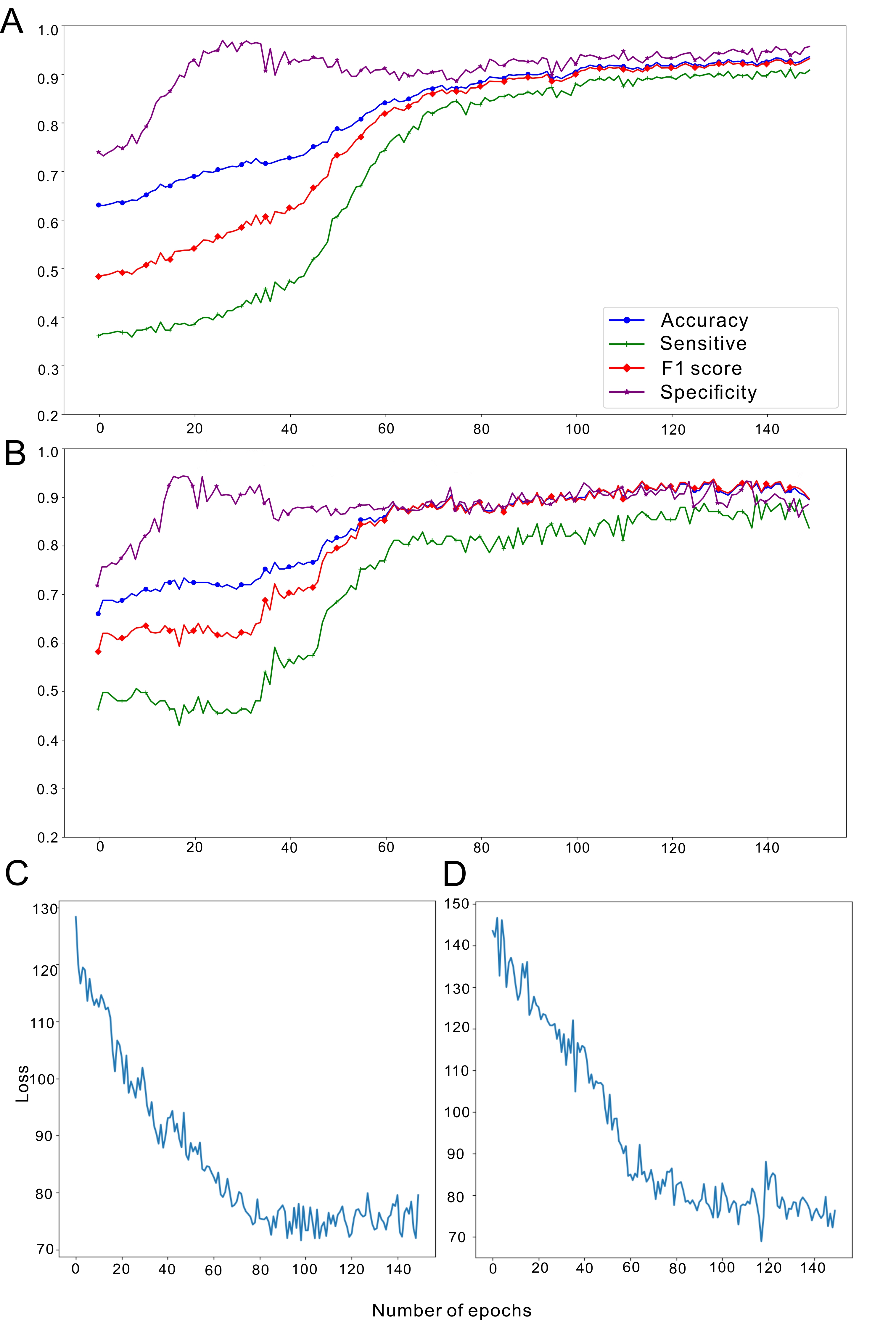}
    \end{center}
    \caption{\textbf{(A)} Curves of relevant parameters in the training process corresponding to the CHB-MIT dataset. \textbf{(B)} Curves of relevant parameters in the training process corresponding to the Siena scalp dataset.\textbf{(C)} The loss value change curve of the CHB-MIT dataset during the training process. \textbf{(D)} The loss value change curve of the Siena scalp dataset during the training process. }\label{train}
    \end{figure}

    The network performance was evaluated using four evaluation criteria. 
    Accuracy measures overall correctness, sensitivity measures correct positive identification, specificity measures correct negative identification, and F1 score considers both precision and recall for accuracy assessment.
    The model parameter curve and loss change curve during training are shown in the Figure \ref{train}, Since the results on the test set can better evaluate the generalization ability of the model on unseen data, we only show the curves of the test set. 
    Once training is finished, the network is deployed onto the neuromorphic chip Xylo.
    This process primarily involves the mapping and quantization techniques mentioned earlier \cite{3.2rock}.
    Afterwards, the original data is imported into the Xylo processor and the model prediction results are observed.
    
\section{Results}
    \subsection{Network performance}
    We visualized the above parameters. Additionally, the loss value was observed during the iteration process, and the curves are shown in the figure.
    Since the results on the test set can better evaluate the generalization ability of the model on unseen data, only the curves of the test set are shown.
    The model parameter curves corresponding to the two data sets are depicted.
    In both datasets, the following measures were obtained: Accuracy of 93.3\% and 92.9\%, sensitivity of 90.4\% and 89.7\%, specificity of 96.7\% and 90.1\%, and F1 score of 91.2\% and 92.3\%.
    However, it is worth noting that when the model is deployed on the Xylo processor, a certain loss in accuracy rate relative to the simulation will occur.
    After the final test on the two data sets, the accuracy can reach 89.87\% and 88.62\%.

    In deep learning, model latency refers to the time interval between when a model receives input data and when it outputs results.
    Model latency is critical for real-time applications \cite{6.2.delay}, in this application, for EEG monitoring of epilepsy patients, the model must quickly detect epilepsy signals and sound an alarm in a short time.
    Therefore, evaluating the latency of the SNN model in this experiment can help determine its usability and practicality in real-time applications, as well as optimize the real-time performance of the model.

\begin{figure}[!htbp]
    \begin{center}
    \includegraphics[width=9cm,scale=.75]{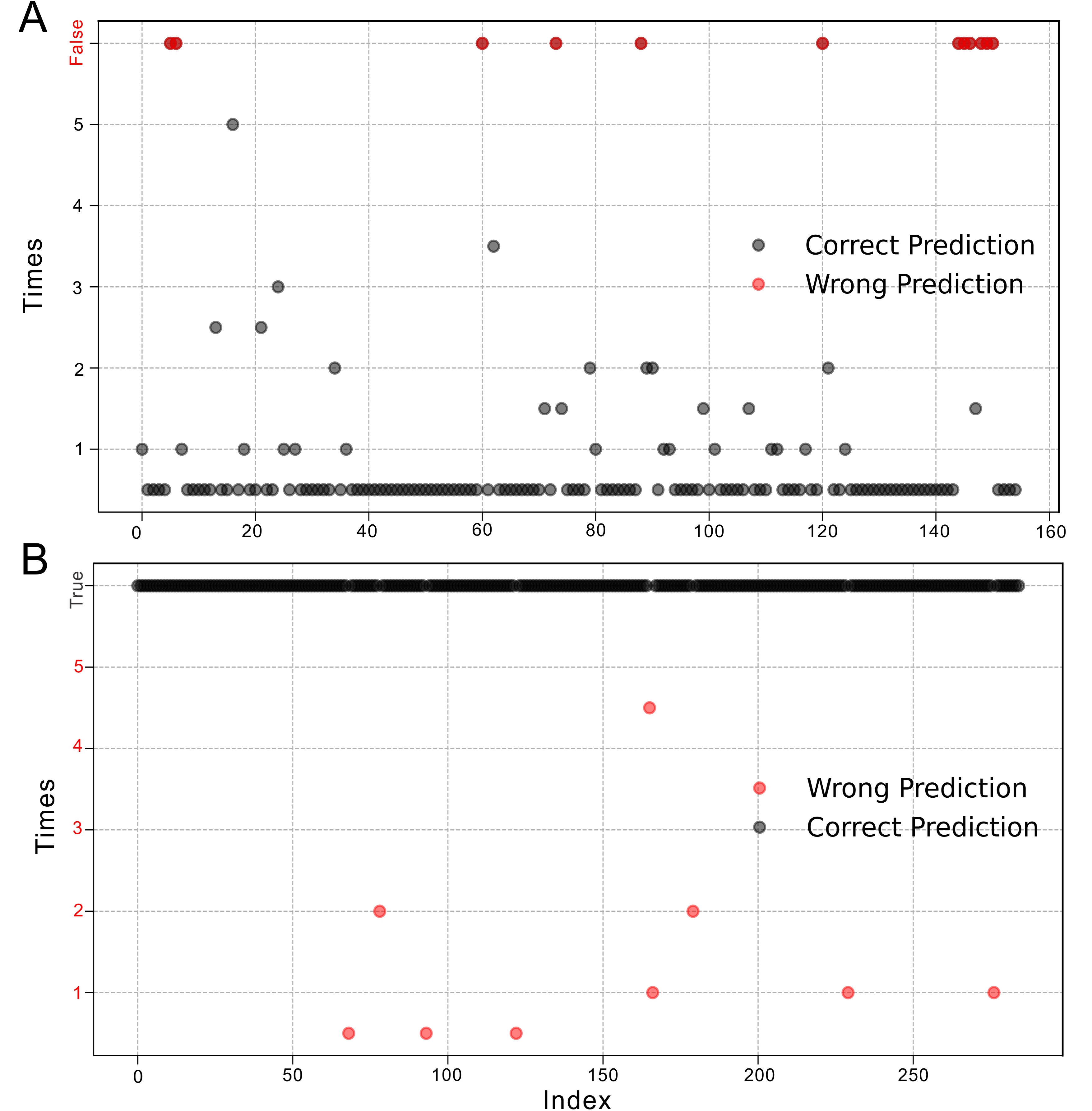}
    \end{center}
    \caption{\textbf{(A)} Latency test results for partially positive samples. The black dot's vertical axis value signifies the accurate prediction time of epilepsy, while the red dot indicates that the model failed to detect epilepsy even after surpassing the corresponding sample time. \textbf{(B)} Latency test results for partial negative samples. The red dot signifies that the model wrongly classified a non-epileptic signal as an epileptic signal at the corresponding time on the y-axis. The black dot denotes that the model accurately classified the signal as non-epileptic even after the sampling time for the sample has passed.}\label{delay}
    \end{figure}

    The SNN was deployed on the Xylo processor, and measurements on delays were performed using the CHB-MIT dataset.
    Each time step was set to 0.5 seconds, and it was found that the majority of 5-second epileptic signals were detected within 0--1 seconds, with a median delay of 0.5 seconds.
    Favorable results were also achieved using the Siena scalp dataset with a delay of 0.75 seconds.
    The visualization of our findings is presented in Figure \ref{delay}.
    
    In order to fully showcase the methods and implementation of our research, we have publicly released the complete source code of the project on GitHub. The code for the project related to the Siena scalp EEG database can be accessed at \url{https://github.com/liruixinxinxin/siena_work}, and for the CHB-MIT scalp EEG database, it is available at \url{https://github.com/liruixinxinxin/epilepsy_complete}. These repositories include detailed documentation of the algorithms and implementation procedures, aimed at facilitating further verification and replication of our research results.
    
    \subsection{Real-time monitoring and power consumption}
\begin{figure*}[hbt!]
    \begin{center}
    \includegraphics[width=16cm,scale=.75]{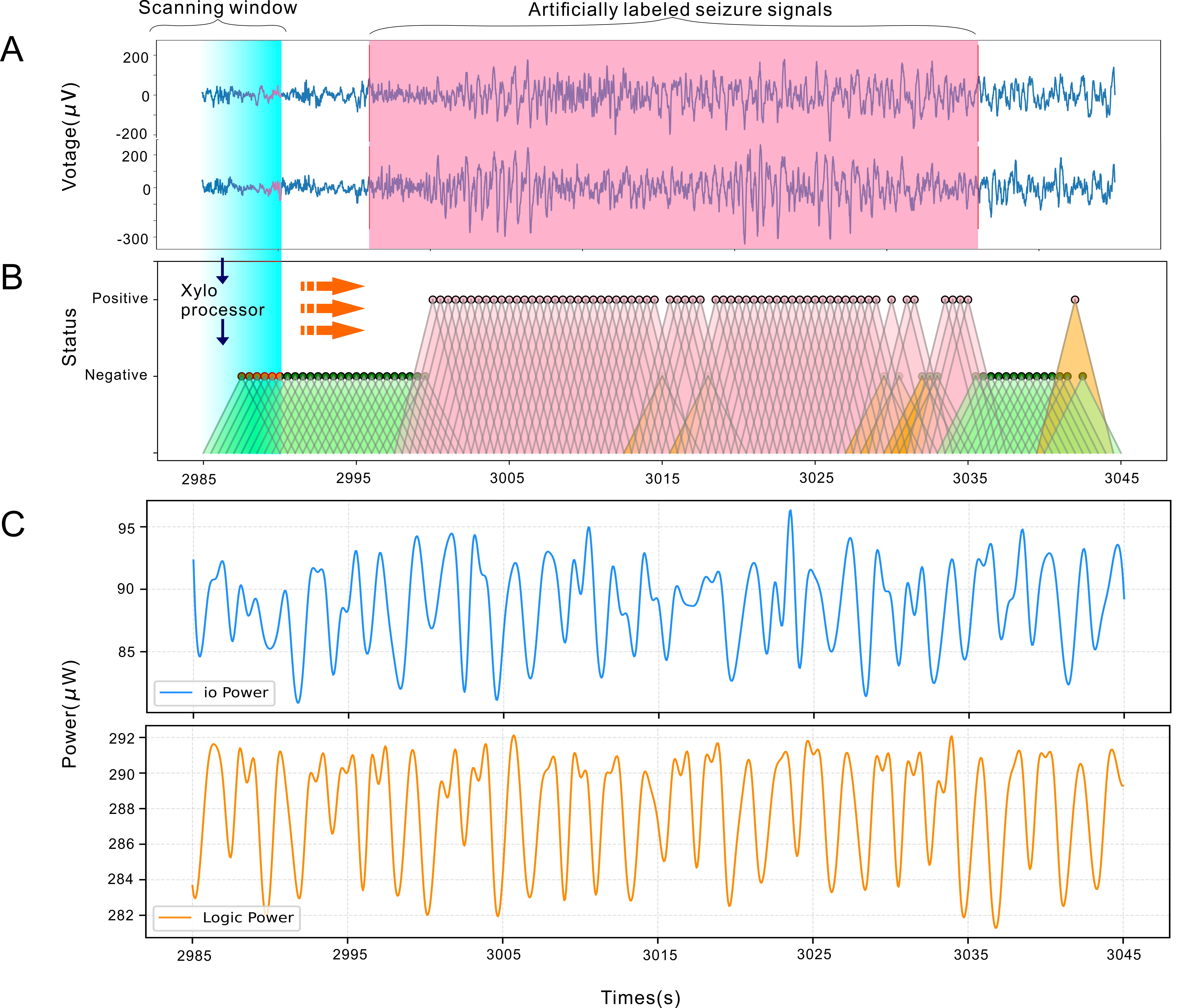}
    \end{center}
    \caption{\textbf{(A)} The official marked epileptic seizure moment is displayed in light red on the initial signal map, while the 5-second scanning window is represented in blue. The application allows for quick diagnosis during real-time signal recording. \textbf{(B)} The time interval corresponding to each vertex of the triangle is the time period in the original signal corresponding to the predicted result. By diagnosing the original signal, the model can identify the silent period in green, the epileptic period in red, and the false touch automatically recognized by the model in yellow. \textbf{(C)} The tracking of power consumption occurs over a period of time.}\label{power}
\end{figure*}

    For power consumption measurement, we enable the power recording function by setting \texttt{"power\_record = True"} in the code. actually calls a function related to the Samna interface, allowing the system to automatically read voltage and current information from the chip's registers. This method, which obtains precise data directly from the hardware, enables researchers to monitor and evaluate the chip's energy consumption in real time under various operating conditions.
    A real-time epilepsy detection experiment was conducted and the power consumption during the monitoring process was measured.
    As an example, The first seizure of patient No. 1 from CHB-MIT was used (show in Fig. \ref{power}).
    The data sample reveals that patient No. 1 experienced an epileptic seizure from 2996 seconds to 3036 seconds.
    The model issued a red alarm at 2997.5 seconds, and ended the alarm at 3033.5 seconds.
    False alarms during this process may occur, which were addressed by carrying out post-processing in the algorithm.
   Following an analysis of the model's accuracy and the trade-off between false alarms and sensitivity, the algorithm triggers or cancels alerts only when four consecutive test values are either 1 or 0. The purpose behind selecting this specific threshold is to minimize false alarms while retaining a high sensitivity to epileptic seizures.
    The device exhibited a consistent average I/O power range of 80.2--95.5~$\mu$W, with an average consumption of 87.4~$\mu$W.
    The digital SNN core and control logic showed stable power consumption within 280.3--292.1~$\mu$W, averaging at 287.9~$\mu$W.
    Our experimental results confirmed that the power consumption of device is in the microwatt range, significantly lower than the power consumption reported in previous research involving traditional on-chip biosignal detection.
    Table \ref{compare of power} displays a comparison of power consumption and other metrics with similar works.
    
\begin{table}[htp]
    \caption{Comparison of chip power consumption with other similar works.$^1$ Multilayer Perceptron. $^2$ Support Vector Machine. $^3$ Long Short-Term Memory. $^4$ Convolutional Neural Network. $^5$ Spiking Neural Network.$^6$ Hilbert Transform.$^7$Discrete Wavelet Transform.}
    \label{compare of power} 
    \centering
    \small
    \resizebox{\textwidth}{!}{
    \begin{tabular}{>{\centering\arraybackslash}p{1.8cm}>{\centering\arraybackslash}p{1.8cm}>{\centering\arraybackslash}p{1.8cm}>{\centering\arraybackslash}p{1.8cm}>{\centering\arraybackslash}p{1.9cm}>{\centering\arraybackslash}p{1.8cm}>{\centering\arraybackslash}p{1.8cm}>{\centering\arraybackslash}p{1.8cm}}
    \hline
    \textbf{Work} & \cite{power1} & \cite{power2} & \cite{power3} & \cite{power4} & \cite{power5} & \cite{power6} & This work \\ \hline
    \multirow{2}{*}{\textbf{Processor}} & M2GL025-VF256 IGLOO2 FPGA by Microsemi & FPGA & Nexys 4 Artix 7 FPGA & M2GL025-VF256 IGLOO2 FPGA & Neuromorphic chip & Xilinx Zedboard FPGA & Neuromorphic processor Xylo \\ \hline
    \textbf{Power(µW)} & 159700 & 1589 & 284000 & 448 & 160000 & 166000 & 287.9 \\ \hline
    \textbf{Network Architecture} & MLP$^1$ and ANN & ANN & SVM$^2$ & SVM$^2$ & LSTM$^3$ and CNN$^4$ & MLP$^1$ and ANN & SNN$^5$ \\ \hline
    \textbf{Technology Used} & / & TSMC 0.18 $\mu$Wm CMOS & / & TSMC 65nm 1P9M CMOS & 55 nm CMOS & / & 28nm CMOS \\ \hline
    \textbf{Classification Method} & Feature extraction using HT$^6$, classification with MLP$^1$ & Feature extraction using HT$^6$, classification with MLP$^1$ & Feature extraction using DWT$^7$, classification with SVM$^2$ & Classification using lifting wavelet transform and SVM$^2$ & Classification using 2D-CNN$^4$ and Bi-LSTM$^3$ & Three-layer MLP$^1$, trained and tested on FPGA & Real-time stream data analysis, parallel processing \\ \hline
    \textbf{Chip Size} & / & 1409.12 x 1402.36 µm² & / & 0.98 mm² & / & / & 6.5 mm² \\ \hline
    \textbf{Power Voltage(V)} & / & Core 1.8 I/O 3.3 & / & 1  & / & Core 1.8, IO 3.3 & Core 1.8, IO 2.5 \\ \hline
    \textbf{Frequency} & / & 8 MHz & / & 1 MHz & / & 100 MHz & 250 MHz \\ \hline
    \textbf{SRAM} & / & 144 x 18 bits & / & 256×32bits & / & / & 124KB \\ \hline
    \textbf{Measurement Method} & Actual measurement & Actual measurement & Theoretical simulation estimate & Actual measurement & Actual measurement & Actual measurement & Real-time actual measurement \\ 
    \hline
    \end{tabular}}
\end{table}

\begin{table*}[bht]
    \caption{Comparison of results on CHB-MIT Scalp EEG and SINEA Scalp EEG datasets. $^1$Short-time Fourier Transform. $^2$Convolutional Neural Network. $^3$ Binary Single-Dimension. $^4$ Recurrent Neural Network. $^5$  Hierarchical Neural Network. $^6$ Support Vector Machine. $^7$ Artificial Neural Network. $^8$ Artificial Neural Network. Acc: Accuracy. Sens: Sensitivity. Spec: Specificity. H/W Deploy: Hardware Deployment}
    \label{compare} 
    \centering
    \begin{minipage}{\linewidth}
    \centering
    \small
    \begin{tabular}{llllllll} 
    \hline
    \textbf{Dataset}& \textbf{Citation} & \textbf{Method} & \textbf{Acc. \%} & \textbf{Sens. \%} & \textbf{Spec. \%} & \textbf{No. weights} &\textbf{H/W Deploy} \\[1ex]
    \multirow{6}{*}{CHB-MIT}
    & \cite{chb1} & STFT$^1$-CNN$^2$ & N.A. & 81.2 & --- & 119K & No \\
    & \cite{chb2} & BSD$^3$-CNN & N.A. & 94.6 & --- & 53.1K & No\\
    & \cite{chb3} & Deep CNN & 90.0 & 90.2 & 88.0 & 70.9K  & No\\

    & \cite{chb6} & CNN & 61.0 & 59.0 & --- & 8.93M & No\\
    & \cite{chb5} & HNN$^5$ & 98.9 & --- & --- & $>$ 27.3M & No\\
    
    & \cite{chb4} & CNN+RNN$^4$ & 96.2 & 98.2 & 94.0 & ---  & No\\
    & \cite{chb7} & CNN & 95.5 & 94.7 & -- & 1.57M  & No\\
    & \cite{chb8} & CNN & 96.9 & 97.0 & 96.7 & 10,854  & No\\
    &\textbf{This work} & \textbf{SNN} & \textbf{93.3} & \textbf{90.4} & \textbf{96.7} &\textbf{2.4K} & \textbf{Yes}\\
    \hline
    \multirow{3}{*}{SINEA}
    & \cite{siena1} & SVM$^6$ & 87.5 & 85.0 & 90.2 & --- & No\\

    & \cite{seina2} & ANN$^7$ & 94.0 & 74.0 &  --- &   --- & No\\
    & \textbf{This work} & \textbf{SNN} &\textbf{92.9}  &\textbf{89.7} &\textbf{90.1} & \textbf{2.4K}& \textbf{Yes}\\
    \hline
    \end{tabular}
    \end{minipage}
\end{table*}

    \subsection{Performance comparison with existing work}
    By conducting several experiments and summarising the results in Table. \ref{compare}, we observe that the overall performance of the SNN model is somewhat lower than that of the traditional artificial neural network.
    However, in this project, the scale of the SNN model is relatively small, which enables the model to better simulate the neural system, more effectively process time series data, and use computing resources more efficiently.
    This is because in deep learning models, the number of weights and the power consumption of the model are usually related.
    The more weights there are, the greater the complexity and computational load of the model, which may require more computing resources and higher power consumption during deployment.
    In addition, the number of weights also affects the model's storage and data transmission requirements, which may also lead to higher power consumption.
    Therefore, using a very small number of weights in this model, the experiment results demonstrate that microwatt-level power loss can be achieved.
    Currently, many related studies on power consumption measurement remain at the theoretical estimation stage\cite{add1.3,add1.1}. In contrast, we have measured and visualized the power values in real-time epilepsy detection. While some studies may achieve similarly low power consumption, they often suffer from significantly higher latency\cite{add1.1,add1.2}. Our study, however, achieved a balanced outcome by ensuring low power consumption, low latency, and high accuracy.

\section{Discussion and future work}
    In this research, a real-time epilepsy detection method based on SNN is proposed and implemented on a hardware platform using the neuromorphic processor Xylo.
    This study presents several advantages and innovations.
    Firstly, a third-generation spiking neural network was utilized.
    In processing EEG signals, compared with ANN, SNN can better simulate the behavior of neurons, especially the process in which neurons fire spikes within a certain time window after receiving inputs. Therefore, SNN can more accurately capture the key features of EEG signals and perform accurate processing.
    Second, Instead of providing complete or buffered input signals into a neural network, our SNN analyses the EEG signals in a real-time streaming mode.
    This working mode can improve the efficiency and low latency of the method, which significantly enhances the clinical usability for EEG-based seizure detection. Compared with buffering the input signal, real-time streaming mode analysis signal does not require storing large amounts of data, which can avoid wasting storage space and delay problems. This is very important for designing low-power and implantable brain-computer interfaces and related devices. In addition, using real-time streaming mode analysis of EEG signals can help SNN better adapt to changes in different real-time EEG data streams, thereby improving the method's adaptability and generalization ability.
    Finally, our network is deployed on the neuromorphic chip Xylo instead of traditional von~Neumann architecture chips, this is no longer the traditional work of simply simulating biological signals in SNN.
    Unlike the traditional von~Neumann architecture, Neuromorphic chips natively use neuron and synaptic models for computation; have strong parallel computing capabilities; and can efficiently process large datasets in streaming mode.
    In addition, mear-memory-compute Neuromorphic processors, which integrate memory closely with computing resources, consume significantly less energy than traditional von~Neumann architectures.
    Our results highlight the broad prospects for application of Neuromorphic chips in the fields of edge machine learning and artificial intelligence inference.
    
    Our method achieves relatively high accuracy in detecting epilepsy, but there is still room for improvement.
    Specifically, the accuracy rate of the proposed method is not be as high as that of a traditional ANN implementation.
    Additionally, the generalization ability of the model can be improved.
    This is a trade-off between accuracy and model simplicity (and low power).
    
    Our proposed network requires the smallest amount of resources while achieving comparable training metrics (Table \ref{compare}). Additionally, in Table \ref{compare of power}, we compared the processor types, network architectures, technologies used, classification methods, chip sizes, power supply voltages, frequencies, SRAM capacities, and measurement methods of different works. It can be seen that the neuromorphic processor Xylo used in this work has the lowest power consumption (287.9 µW), which is the best among similar works.

    Our proposed solution can assist medical practitioners in accurately diagnosing patients with epilepsy, thereby facilitating timely and effective treatment plans and enhancing the quality of life of such patients.
    Furthermore, our work paves the way for the development of low-power wearable devices for biosignal detection, and presents novel insights and tools for neurological disease diagnosis and treatment, as well as other clinical medicine fields in brain science research.
   In addition, this method can be extended to process other bio-signals, such as the classification of electromyography (EMG) and electrocardiography (ECG). The use of real-time streaming signal analysis can provide more efficient and reliable solutions for the detection and treatment of muscle and heart diseases\cite{ECG}. Therefore, the potential applications of SNN in the medical domain are extensive, and they hold considerable research and practical significance.

\section*{Declaration of competing interest}
    The authors of this paper hereby declare that there are no financial, personal, or other relationships that could or might be perceived to influence the objectivity of our research and the results presented in this paper. We affirm that our research work has been conducted independently and has not been unduly influenced by any funding agencies, commercial entities, or other organizations.

\section*{Acknowledgments}

    This work was financially supported by the National Natural Science Foundation of China (32260244), Hainan Provincial Natural Science Foundation of China (524YXQN\\416, 824CXTD424), Science and Technology Special Fund of Hainan Province (ZDYF2022SHFZ289) and the Research Foundation of One Health Collaborative Innovation Center of Hainan University (XTCX2022JKC01). 

\bibliographystyle{unsrt}  
\bibliography{references}

\end{document}